\documentstyle[prd,aps,floats,epsfig]{revtex}
\begin{document}
\draft
%@twocolumnfalse\endcsname 

\title{Geodesic Deviation in Kaluza-Klein Theories\thanks{NIKHEF/00-029}}
\author{Richard Kerner$^1$\thanks{\tt rk@ccr.jussieu.fr}, 
J\'er\^ome Martin$^2$\thanks{\tt jmartin@iap.fr},
Salvatore Mignemi$^{3}$\thanks{\tt smignemi@vaxca1.unica.it}  
and Jan-Willem van Holten$^4$\thanks{\tt
v.holten@nikhef.nl} }
\address{$^1$ LGCR, Universit\'e Paris-VI, Tour 22,
Bo\^{\i}te 142, 4, Place Jussieu, 75005 Paris, France.
\\
$^2$ Institut d'Astrophysique de Paris, 98 boulevard Arago, 
75014 Paris, France. 
\\
$^3$ Dipartimento di Matematica, Universit\`a 
di Cagliari, viale Merello 92, 09123 Cagliari, Italy. 
\\
$^4$ NIKHEF, POBox 41882, 1009 DB Amsterdam, The Netherlands.}

\maketitle

\begin{abstract} 
We study in detail the equations of the geodesic deviation in
multidimensional
theories of Kaluza-Klein type. We show that their $4$-dimensional
space-time
projections are identical with the equations obtained by direct variation
of the usual geodesic equation in the presence of the Lorentz force,
provided
that the fifth component of the deviation vector satisfies an extra
constraint derived here.
\end{abstract}

\vskip1cm

\section{Introduction}

It is well known that the equation satisfied by a world-line $x^{\mu}(\tau
)$ 
of a massive charged particle in a $4$-dimensional Riemannian space-time
in
presence of an electromagnetic field is given by
\begin{equation}
\label{geo4}
\frac{{\rm d}^2 x^{\mu}}{{\rm d} \tau^2} + \Gamma ^{\mu }_{\nu \lambda}
\frac{{\rm d} x^{\nu}}{{\rm d} \tau}\frac{{\rm d} x^{\lambda}}{{\rm d}
\tau}  
+ \frac{q}{m}F_{\lambda }{}^{\mu}\frac{{\rm d} x^{\lambda}}{{\rm d}
\tau}=0,
\quad \mu, \nu,\dots = 0,\dots ,3.
\end{equation}
In this expression, $m$ and $q$ are the mass and the charge, respectively,
of a test particle, $\tau $ is the $4$-dimensional proper time and the 
electromagnetic field $F_{\mu \nu }$ is defined by $F_{\mu \nu }\equiv 
\partial _{\mu }A_{\nu }-\partial _{\nu }A_{\mu}$ where $A_{\mu}
(x^{\kappa })$ is the vector potential. The $\Gamma ^{\mu }_{\nu \lambda}$
are the $4$-dimensional Christoffel symbols. The second term of the above
equation represents the inertial force whereas the last term is the
Lorentz
force.
\par
The study of geodesics in multidimensional theories of Kaluza-Klein type
has
been performed by many authors \cite{Ker1} -  \cite{Arodz2} in quite an
exhaustive manner. It has been known since the very beginning \cite{Kaluza,Klein}
that, in the simplest version of the theory (without scalar field), the
geodesic equation in $5$ dimensions coincides with the geodesic equation
in
$4$ dimensions with an extra term which can be identified with the Lorentz
force. Indeed, in $5$ dimensions, the geodesic equation is given by
\begin{equation}
\frac{{\rm d}^2 x^A}{{\rm d} s^2} \, + {\displaystyle{A \brace{BC}}} \, 
\frac{{\rm d} x^B}{{\rm d} s} \frac{{\rm d} x^C}{{\rm d}s} = 0 , \,
\ \ \ \ \, \ \ \, \ \  \, A,B,\dots = 1,\dots ,5,
\end{equation}
where $s$ is the $5$-dimensional interval length, and the brackets
${\displaystyle{A \brace{BC}}}$ are the $5$-dimensional Christoffel
symbols.
Projecting this equation under the assumption that all the quantities are
independent of the fifth coordinate, $\partial_5 = 0$, it is easy to
establish, using the formulae displayed in the appendix, that one obtains
\begin{equation}
\frac{{\rm d}^2 x^{\mu}}{{\rm d} s^2} + \Gamma ^{\mu }_{\nu \lambda}
\frac{{\rm d} x^{\nu}}{{\rm d} s}\frac{{\rm d} x^{\lambda}}{{\rm d} s}  
+\biggl(\frac{{\rm d}x^5}{{\rm d}s} + A_{\mu} 
\frac{{\rm d}x^{\mu}}{{\rm d}s}\biggr)F_{\lambda }{}^{\mu}
\frac{{\rm d} x^{\lambda}}{{\rm d} s}=0, \quad
\frac{{\rm d}}{{\rm d}s}\biggl(\frac{{\rm d}x^5}{{\rm d}s} + A_{\mu} 
\frac{{\rm d}x^{\mu}}{{\rm d}s}\biggr)=0. 
\label{eom}
\end{equation} 
The second equation, telling us that the quantity $Q \equiv
\frac{{\rm d}x^5}{{\rm d}s} + A_{\mu} \frac{{\rm d}x^{\mu}}{{\rm d}s}$ is
constant along the $5$-dimensional geodesics parametrized by $s$, reflects 
the fact that the backgrounds have been chosen such that $x^5$ is a cyclic 
co-ordinate. Since we have the following relation between the squares of
the 
intervals in $5$ and $4$ dimensions:
\begin{equation}
{\rm d} s^2 = (g_{\mu \nu} + A_{\mu} A_{\nu}) \,
{\rm d} x^{\mu} {\rm d} x^{\nu} + 2 \, A_{\mu} {\rm d} x^{\mu} {\rm d} x^5 
+ ({\rm d} x^5)^2={\rm d} \tau^2 + \biggl(\frac{{\rm d} x^5}{{\rm d} s} 
+ A_{\mu} \frac{{\rm d} x^{\mu}}{{\rm d} s} \biggr) 
\biggl(\frac{{\rm d} x^5}{{\rm d} s} 
+ A_{\nu} \frac{{\rm d} x^{\nu}}{{\rm d} s} \biggr) \, {\rm d} s^2,
\end{equation}
which amounts to ${\rm d}s^2(1 - Q^2) = {\rm d} \tau^2$, we see that the 
$4$-dimensional equation (\ref{geo4}) is recovered provided that we 
make the following identification :
\begin{equation}
\frac{q}{m}=\frac{Q}{\sqrt{1-Q^2}} \hspace{1em} \Leftrightarrow 
 \hspace{1em} Q = \frac{q/m}{\sqrt{1 + (q/m)^2}}.
\label{identify}
\end{equation}
This means that we suppose that $Q^2 < 1$, so that ${\rm d}s$ is timelike 
whenever ${\rm d} \tau$ is timelike, and vice versa. A more general
situation is discussed in detail in \cite{Leibowitz}. The non-abelian
generalization has been considered in \cite{Ker1} and \cite{Wong}. Since
the multidimensional theories of Kaluza-Klein type are constructed as
copies
of Einstein's General Relativity theory in more than $4$ dimensions, all
the
usual mathematical corollaries remain valid. For example, it is possible
to
cancel the Christoffel symbols along a given geodesic curve by an
appropriate
choice of coordinates, which amounts to the annulation of forces acting on
a test particle moving along that geodesic line. In $4$ dimensions, the
cancellation along the worldline of the $5$-dimensional Christoffel
symbols
may be interpreted as the simultaneous compensation of gravitational and
the
Lorentz forces by an appropriate acceleration field.
\par
It is also a well known fact that such cancellation can be performed along
only one given geodesic line at once, but not necessarily along all its
neighbors. This fact becomes obvious when one looks at the {\it geodesic
deviation} equation
\begin{equation}
\label{devi4}
\frac{{\rm D}^2 (\delta x^{\mu})}{{\rm D} \tau^2}= 
{}^4R^{\mu}{}_{\rho \nu \lambda}\frac{{\rm d} x^{\rho }}{{\rm d} \tau} 
\frac{{\rm d} x^{\nu}}{{\rm d} \tau}\delta x^{\lambda } .
\end{equation}
where $\delta x^{\lambda }$ is an infinitesimal ``geodesic deviation''
vector,
and ${\rm D}/{\rm D}\tau $ denotes the pull-back of covariant derivatives
along the time-like geodesics. For a massive {\it and} charged test
particle
in the presence of both gravitational and electromagnetic fields
($R^{\mu}_{\, \ \ \nu \lambda \rho} \neq 0$ and $F_{\mu \nu} \neq 0$), it
is not difficult to derive the generalized world-line deviation equation
by taking direct variation of the world-line equation (\ref{geo4}):
\begin{equation}
\label{devi4F}
\frac{{\rm D}^2 (\delta x^{\mu})}{{\rm D} \tau^2}= 
{}^4R^{\mu}{}_{\rho \nu \lambda}\frac{{\rm d} x^{\rho }}{{\rm d} \tau} 
\frac{{\rm d} x^{\nu}}{{\rm d} \tau}\delta x^{\lambda }+
\frac{q}{m} \, \biggl[ (\nabla_{\rho} F^{\mu}_{\, \ \ \nu}) 
\frac{{\rm d} x^{\nu}}{{\rm d} \tau} \delta x^{\rho} +  F^{\mu}_{\, \ \
\nu}
\frac{{\rm D}(\delta x^{\nu})}{{\rm D} \tau} \biggr],
\end{equation}
Here the Riemann tensor appears explicitly, making it automatically
impossible to cancel its influence by any local or global coordinate or
gauge transformations. This is why the study of the geodesic deviation
in multidimensional theories, which reads
\begin{equation}
\label{devi5}
\frac{{\rm D}^2(\delta x^A)}{{\rm D} s^2}=R^A{}_{BCE}
\frac{{\rm d}x^B}{{\rm d} s} \frac{{\rm d} x^C}{{\rm d} s} \delta x^E, 
\end{equation}
is of particular interest. Indeed, when explicited in the form that splits
up the $4$-dimensional space-time and the $D$-dimensional internal space,
{\it a priori}, new terms show up, containing quadratic expressions of the
type $F^{\mu}_{\, \ \ \nu}F_{\mu \lambda}$ appearing in the
$5$-dimensional
Riemann tensor (cf. Appendix), that can not be foreseen or derived
from a purely $4$-dimensional point of view, even if one tries to
introduce
the interaction of charges with gauge and scalar fields. It will fix in a
canonical way the terms describing the purely gravitational influence of 
those fields, which by their energy density must influence the
trajectories
of chargeless massive particles, too, provoking tidal effects which should
deform  the initially parallel geodesic lines.
\newline

%%Begin InstantTeX Picture
\let\picnaturalsize=N
\def\picsize{2.5in}
\def\picfilename{diag.eps}
%If you do not have the picture file add:
%\let\nopictures=Y
%to the beginning of the file.
\ifx\nopictures Y\else{\ifx\epsfloaded Y\else\input epsf \fi
\let\epsfloaded=Y
\centerline{\ifx\picnaturalsize N\epsfxsize \picsize\fi
\epsfbox{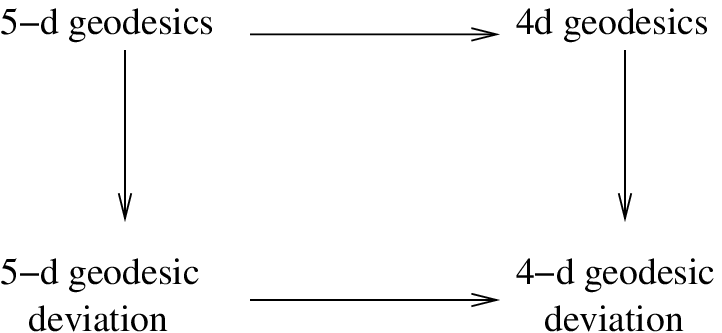}}}\fi
%%End InstantTeX Picture 
\begin{center} {\footnotesize{Fig.1: Kaluza-Klein reduction of geodesic
               deviations}} \end{center} 
\vspace{1ex} 

\noindent
Hence the question arises whether calculation of the geodesic deviation 
after reduction to 4 dimensions yields the same result as projecting 
the 5-dimensonal geodesic deviations to 4 dimensions, rendering the 
diagram of Fig.\ 1 commutative. This is of course an important issue 
since any difference could be used to discriminate a purely
$4$-dimensional 
theory from a Kaluza-Klein approach. 
\par
The purpose of this short technical note is to address this question. To
our 
knowledge, the analysis of this problem can not be found in the existing
literature, and we believe that the present study will close this gap, and
in addition, will shed some new light on the interplay between the gauge
fields and gravitation, and on the interpretation of the equivalence
principle in multidimensional theories as well.

\section{Geodesic deviation equations in $4+1$ dimensions}

We consider here the version of the $5$-dimensional Kaluza-Klein theory in
which the scalar field $\varphi$ is put equal to $1$ from the beginning,
which
makes the field equations arising from the variational principle in $5$
dimensions strictly equivalent to the Einstein-Maxwell system. We shall 
perform all the calculations in a holonomous coordinate system in order to
make the interpretation of the geodesic equations and the affine
parameters
as straightforward as possible. The space-time components obey the
equation
\begin{eqnarray}
\frac{{\rm D}^2 (\delta x^{\mu})}{{\rm D}s^2} &=& [{}^{4}R^{\mu}{}_{\nu
\rho \kappa} - \frac{3}{4} F^{\mu}{}_{\rho }F_{\nu \kappa } ] 
\frac{{\rm d} x^{\nu}}{{\rm d} s} \frac{{\rm d} x^{\rho }}{{\rm d} s}  
\delta x^{\kappa } - \frac{Q}{2}[\nabla_{\kappa } F_{\rho}{}^{\mu } + 
\nabla ^{\mu } F_{\rho \kappa}]\frac{{\rm d} x^{\rho }}{{\rm d} s} 
\delta x^{\kappa } - \frac{1}{2}(A_{\kappa}\delta x^{\kappa } + \delta
x^5)
(\nabla_{\rho} F^{\mu}{}_{\nu}) \frac{{\rm d} x^{\rho }}{{\rm d} s}
\frac{ {\rm d} x^{\nu }}{{\rm d} s}
\nonumber \\
& & -\frac{Q^2}{4}F^{\mu}{}_{\lambda}F_{\rho }{}^{\lambda}\delta x^{\rho }
+\frac{Q}{4}(A_{\kappa}\delta x^{\kappa }+\delta x^5)
F^{\mu }{}_{\lambda}F_{\rho}{}^{\lambda} \frac{{\rm d} x^{\rho }}{{\rm d}
s} 
\label{geomu}
\end{eqnarray}
The equation (\ref{geomu}) still is not explicit enough to be solved as 
function of the space-time variables. The derivatives with respect to 
$5$-dimensional interval ${\rm d}s$ should be replaced by the derivatives 
with respect to the $4$-dimensional proper time ${\rm d}\tau$. Next, the
second-order covariant derivative appearing on the right-hand side and
containing the $5$-dimensional connection coefficients and their partial
derivatives, has to be expressed in terms of ordinary derivatives and
$4$-dimensional Christoffel symbols along with the gauge-invariant
quantities
containing the Faraday tensor $F_{\mu \nu}$. The relation between the
covariant derivations with respect to the parameters ${\rm d}s$ (the
$5$-dimensional line element) and ${\rm d}\tau$ (particle's proper time in
$4$ space-time dimensions) is quite complicated when it comes to the
second-order covariant derivatives. This is why we omit all the
intermediary
calculations, giving here the final result. The second covariant
derivatives
are related as follows for the $4$-dimensional components of a given
$5$-dimensional vector $u^A$:
\begin{eqnarray}
\label{der54}
\biggl(\frac{{\rm d}s}{{\rm d}\tau}\biggr)^2\frac{{\rm D}^2 u^{\mu }}{{\rm
D} s^2} &=& \frac{{\rm D}^2 u^{\mu  }}{{\rm D}\tau^2} 
+ \frac{1}{2}\frac{Q}{\sqrt{1-Q^2}}\nabla_{\rho }F_{\nu }{}^{\mu }
\frac{{\rm d}x^{\rho }}{{\rm d}\tau }u^{\nu } 
- \frac{1}{2}(A_{\kappa}u^{\kappa }+u^5)\nabla_{\rho }F_{\, \ \ \nu }^{\mu
}
\frac{{\rm d}x^{\rho }}{{\rm d}\tau }\frac{{\rm d}x^{\nu }}{{\rm d}\tau }
+\frac{Q}{\sqrt{1-Q^2}}F_{\nu }{}^{\mu }\frac{{\rm D}u^{\nu }}{{\rm
D}\tau}
\nonumber \\
& &+\frac{1}{4} \, \frac{Q^2}{(1-Q^2)} \, F_{\nu}{}^{\mu }
F_{\rho }{}^{\nu}u^{\rho } - \frac{1}{4} \, \frac{Q}{\sqrt{1-Q^2}}
(A_{\kappa}u^{\kappa }+u^5)F_{\nu}{}^{\mu }
F_{\rho}{}^{\nu }\frac{{\rm d}x^{\rho }}{{\rm d}\tau }
- \frac{3}{4} \, F_{\, \ \ \lambda }^{\mu }F_{\kappa \rho }  
\frac{{\rm d} x^{\lambda }}{{\rm d} \tau } \frac{{\rm d} x^{\kappa }}{{\rm
d} \tau}u^{\rho }
\nonumber \\
& &+F_{\lambda }{}^{\mu }\frac{{\rm d} x^{\lambda }}{{\rm d} \tau }
\biggl[\frac{{\rm d}}{{\rm d}\tau} (A_{\kappa}u^{\kappa }+u^5) +
F_{\kappa \rho } \frac{{\rm d} x^{\kappa}}{{\rm d} \tau } \, u^{\rho
}\biggr].
\end{eqnarray}
Combining equations (\ref{geomu}) and (\ref{der54}) and using homogeneous
Maxwell's equations, $\nabla_{\lambda} F^{\mu}{}_{\nu} + 
\nabla_{\nu} F_{\lambda}{}^{\mu} + \nabla^{\mu} F_{\nu \lambda} = 0$
and the identification of the physical charge-to-mass ratio
(\ref{identify}), one obtains
\begin{equation}
\label{inter}
\frac{{\rm D}^2 (\delta x^{\mu})}{{\rm D} \tau^2}= 
{}^4R^{\mu}{}_{\rho \nu \lambda}\frac{{\rm d} x^{\rho }}{{\rm d} \tau} 
\frac{{\rm d} x^{\nu}}{{\rm d} \tau}\delta x^{\lambda }+
\frac{q}{m} \, \biggl[ (\nabla_{\rho} F^{\mu}_{\, \ \ \nu}) 
\frac{{\rm d} x^{\nu}}{{\rm d} \tau} \delta x^{\rho} +  F^{\mu}_{\, \ \
\nu}
\frac{{\rm D}(\delta x^{\nu})}{{\rm D} \tau} \biggr]
+F^{\mu }{}_{\lambda }\frac{{\rm d} x^{\lambda }}{{\rm d} \tau }
\biggl[\frac{{\rm d}}{{\rm d}\tau} (A_{\kappa}\delta x^{\kappa }
+\delta x^5)+F_{\kappa \rho }\frac{{\rm d} x^{\kappa }}{{\rm d} \tau }
\delta x^{\rho }\biggr].
\end{equation}
This equation would coincide with the usual $4$-dimensional deviation
equation
(\ref{devi4F}) if it were not for the last term, which contains the
standard
Lorentz force multiplied by the expression in square brackets, linear in
the infinitesimal deviation vector. However, it is easily recognized that 
the last term just represents the deviation of the 5th component 
of the momentum $Q$, which by equation (\ref{eom}) is conserved, and which 
we identified with the charge through equation (\ref{identify}): 
\begin{equation}
\label{1stintegral}
\delta Q\, =\, \biggl[\frac{{\rm d}}{{\rm d}s}\biggl(\delta x^5 +
A_{\lambda}
\delta x^{\lambda}\biggr) + F_{\lambda \rho}
\frac{{\rm d} x^{\rho }}{{\rm d}s}\delta x^{\lambda}\biggr]. 
\end{equation} 
It is noteworthy that the same result can be obtained by using the
simpler,
but non-covariant form of the geodesic deviation equation (\ref{devi5}):
\begin{equation}
\frac{{\rm d}^2 (\delta x^A)}{{\rm d} s^2} + 2{\displaystyle{A
\brace{BC}}}
\frac{{\rm d} x^B}{{\rm d}s} \frac{{\rm d}(\delta x^C)}{{\rm d}s} 
+\biggl(\partial_D {\displaystyle{A \brace{BC}}}\biggr)\frac{{\rm d}
x^B}{{\rm d}s}
\frac{{\rm d} x^C}{{\rm d}s}\delta x^D = 0, 
\label{noncov5}
\end{equation}
which makes the calculations much less tedious. The fifth component 
of the previous equation leads to
\begin{equation}
\label{vanis}
\frac{{\rm d}}{{\rm d}s}\biggl[\frac{{\rm d}}{{\rm d}s}(\delta x^5 +
A_{\lambda}\delta x^{\lambda}) + F_{\lambda \rho}
\frac{{\rm d} x^{\rho }}{{\rm d}s}\delta x^{\lambda}\biggr]  = 0.
\end{equation}
which means that $\delta Q$ is indeed a constant. From a purely 
mathematical point of view, this 
constant can take on any real value, depending on the arbitrary 
choice of {\it initial conditions}, which include the initial values 
of {\it ten} variables, $\delta x^A (0)$
and $[{\rm d} (\delta x^B ) / {\rm d} s ] (0)$, as it is the case for any
system of five ordinary differential equations of second order. The 
fact that $\delta Q$ is a constant means that 
not all the initial data can be independent. As a matter of fact, the
first derivative of the fifth component of the deviation, 
$[{\rm d} (\delta x^5 ) / {\rm d} s ] (0)$, is an imposed
function of the four-dimensional initial deviations, namely
\begin{equation}
\label{cond5}
\frac{{\rm d}(\delta x^5)}{{\rm d}s}(0)=\delta Q-\biggl[
\frac{{\rm d}}{{\rm d}s}(A_{\lambda}\delta x^{\lambda})
+ F_{\lambda \rho}
\frac{{\rm d} x^{\rho }}{{\rm d}s}\delta x^{\lambda }\bigg](0).
\end{equation}
Requiring that $\delta Q=0$ is the condition which must be imposed if 
we want to maintain a one-to-one
correspondence between the geodesic deviation equation in $5$-dimensional
Kaluza-Klein space and the usual deviation equation in presence of the
electromagnetic field in $4$ dimensions. 
\newline
\indent        
Now, returning to equation (\ref{inter}), we get the final result that can
be stated very simply as follows : 
\vskip 0.2cm
{\it The space-time projection of the five-dimensional Kaluza-Klein
geodesic
deviation equation yields for fixed $Q$ the usual four-dimensional
world-line 
deviation equation in the presence of both gravitational and
electromagnetic 
fields, for particles of the same $q/m$; geodesic deviations between 
five-dimensional worldlines with different values of $Q$ describe the 
four-dimensional deviation of world-lines for particles with different 
values of $q/m$.} 
\vskip 0.2cm
\indent
The only influence of the electromagnetic fields on chargeless particles
comes through the term linear in the $4$-dimensional Riemann tensor, which
is a solution of the coupled Einstein-Maxwell equations (see Ref.
\cite{BvHK}).

\section{Final remarks}

Although the final answer to the problem of the projection of the geodesic
deviation equation from the 5-dimensional Kaluza-Klein metric space onto
its 4-dimensional space-time basis is very simple and does not bring any
surprise, it is worth to be checked (the above calculations have never
been published elsewhere, at least to our knowledge), and does not seem to
be totally trivial. It can be interpreted as a strong {\it equivalence
principle} generalized to the $5$-dimensional theory incorporating 
electromagnetism into geometry.
\par
Our result can be easily generalized to the non-abelian case
\cite{Ker1,Wong}, \cite{Ker2,Balak1}, where the conservation of charge 
$Q$ is replaced by a condition on the rotation of the charge isovector in 
the Lie algebra space in which it takes its value. There is no guarantee 
that the higher-order deviation equations, obtained with a similar 
technique of new independent variation of Einstein-Maxwell equations do 
also project properly, although such a statement seems very plausible.
\par
However, the introduction of the {\it dilaton field}, i.e. supposing that
the radius of the compactified 5-th dimension depends on the space-time
position $x^{\mu}$, may bring new effects leading to certain
anomalies in the deviation equation and its projection onto the usual
space-time. One can apply a similar technique of probing the deviations to
the equations of motions of the $p-branes$ embedded in multi-dimensional
spaces, which represent a natural generalization of geodesic curves in
Kaluza-Klein theories.
\newline
\indent
These developments should become the object of another independent study.

\vspace{0.4cm}
\centerline{\bf Appendix}
\vspace{0.3cm}

The $5$-dimensional metric tensor of the theory reads ($A,B,\dots
=1,2,\dots ,5$):
\begin{equation}
\gamma_{AB} = \pmatrix{g_{\mu \nu} + A_{\mu} A_{\nu} & A_{\nu} \cr
A_{\mu} & 1} , \, {\rm \, \ \ with \, } \ \ {\rm d} s^2 
= \gamma_{AB} \, {\rm d} x^A {\rm d} x^B, 
\label{5metric}
\end{equation}
where $g_{\mu \nu} = g_{\mu \nu} (x^{\lambda})$ and $A_{\mu} = A_{\mu} 
(x^{\lambda})$, which means that we consider that both $g_{\mu \nu}$ and 
$A_{\mu}$ do not depend on the fifth coordinate $x^{5}$. Here are the 
Christoffel symbols of the metric (\ref{5metric}):
\begin{eqnarray}
\label{gammas2}
{\displaystyle{\mu \brace {\nu \lambda}}} &=& \Gamma^{\mu}_{\nu \lambda} +
\frac{1}{2} (A_{\lambda} F_{\nu}^{\, \ \ \mu} + A_{\nu}
F^{\, \ \ \mu}_{\,\lambda}), \quad 
{\displaystyle{5 \brace {\mu \nu }}} = \frac{1}{2} \, ( \nabla_{\mu} \, 
A_{\nu} + \nabla_{\nu} \, A_{\mu}) - \frac{1}{2} \, A^{\rho} (A_{\nu}
F_{\mu \rho} + A_{\mu} F_{\nu \rho}), \\ 
\label{gammas3}
{\displaystyle{\mu \brace{5 \nu}}} &=& {\displaystyle{\mu \brace{\nu 5}}}
= \frac{1}{2} \, F_{\nu}^{\, \ \ \mu}, \quad 
{\displaystyle{5 \brace{5 \mu}}} = {\displaystyle{5 \brace{\mu 5}}} =
- \frac{1}{2} \, A^{\kappa} \, F_{\mu \kappa}, \quad 
{\displaystyle{\mu \brace{5 5}}} = 0, \quad 
{\displaystyle{5 \brace{55}}} = 0. 
\end{eqnarray}
With this in mind, we can proceed with the computation of the components
of
$5$-dimensional Riemann tensor in a holonomous system. Using the
convention :
\begin{equation}
R^A{}_{BCD} = \partial_C {\displaystyle{A \brace{DB}}} - \partial_D 
{\displaystyle{A \brace{CB}}} + {\displaystyle{A \brace{CE}}}
{\displaystyle{E \brace{DB}}} - {\displaystyle{A \brace{DE}}}
{\displaystyle{E \brace{CB}}},
\end{equation}
and after some calculus, using also the Bianchi identities satisfied by
the
tensor $F_{\mu \nu}$, we get:
\begin{eqnarray}
R^{\rho}{}_{\lambda \mu \nu} &=&  {}^{(4)} R^{\rho}{}_{\lambda \mu \nu}
\, + \, \frac{1}{4} \, ( F_{\mu}^{ \ \ \rho} F_{\lambda \nu} \, - \, 
F_{\nu}^{ \ \ \rho} F_{\lambda \mu} + 2 \, F_{\lambda}^{ \ \ \rho}
F_{\mu \nu} ) 
\, - \, \frac{1}{2} \, A_{\lambda} \, \nabla^{\rho} \, F_{\mu \nu} 
+\frac{1}{2} \, (A_{\nu} \nabla_{\mu} 
F_{\lambda}^{ \ \ \rho} - A_{\mu} \nabla_{\nu} F_{\lambda}^{ \ \ \rho})
\nonumber \\
& &+\frac{1}{4} \, A_{\lambda} F_{\sigma}^{ \ \ \rho} ( A_{\mu} 
F_{\nu}^{ \ \ \sigma} - A_{\nu} F_{\mu}^{ \ \ \sigma} ) \, , \\
R^5{}_{\mu \nu \lambda} &=& - {}^{(4)} R^{\rho}{}_{\mu \nu \lambda} 
\, A_{\rho} + \frac{1}{2} \nabla_{\mu} F_{ \nu \lambda} + \frac{1}{4} \,
F^{\ \ \rho}_{\mu} \,( A_{\lambda} F_{\rho \nu} - A_{\nu} F_{\rho
\lambda})
- \frac{1}{2} A_{\mu} A_{\rho} \nabla^{\rho} F_{\lambda \nu} \nonumber \\ 
& &+ \frac{1}{2} \, A_{\rho} (A_{\nu} \nabla_{\lambda} F^{\; \rho}_{\mu} - 
A_{\lambda} \nabla_{\nu} F^{\, \rho}_{\mu})  + \frac{1}{4}
A_{\mu}A^{\sigma}F_{\rho \sigma}(A_{\lambda} F_{\nu}^{ \ \ \rho}
- A_{\nu}F_{\lambda}^{ \ \ \rho}) + \frac{1}{4} A^{\rho} ( F_{\lambda
\rho}
F_{\mu \nu} + F_{\nu \rho} F_{\lambda \mu}+2 F_{\mu \rho}F_{\lambda \nu})
\, ,
\\ R^{\rho}{}_{5 \mu \nu} &=& - \frac{1}{2} \nabla^{\rho} F_{\mu \nu} +
\frac{1}{4} F_{\sigma}^{ \ \ \rho} \, (A_{\mu} F_{\nu}^{ \ \ \sigma} -
A_{\nu} F_{\mu}^{ \ \ \sigma} ) \, , \\
R^{\rho}{}_{\mu 5 \nu} &=& - \frac{1}{2} \nabla_{\nu} F^{\, \rho}_{\mu} +
\frac{1}{4} A_{\mu} F^{\ \ \rho}_{\sigma} F^{\ \ \sigma}_{\nu} , \\
R^5{}_{5 \mu \nu} &=& \frac{1}{4} \, F_{\rho \sigma} A^{\sigma} \, 
( F_{\mu}^{ \ \ \rho} A_{\nu} - F_{\nu}^{ \ \ \rho} A_{\mu} ) +
\frac{1}{2} \, A_{\rho} \nabla^{\rho} F_{\mu \nu} \, ,\\
R^5{}_{\mu 5 \nu} &=& \frac{1}{4} \, F_{\mu}^{\ \ \rho} F_{\nu \rho} + 
\frac{1}{2} \, A_{\rho} \nabla_{\nu} F_{\mu}^{\ \ \rho} - \frac{1}{4} \,
A_{\mu} A^{\lambda} F_{\nu}^{\ \ \sigma} F_{\sigma \lambda} \, ,\\
R^{\rho}{}_{5 \lambda 5} &=& - \frac{1}{4} F^{ \ \ \rho}_{\sigma}
F^{\ \ \sigma}_{\lambda} \, , \\  R^5{}_{5 \lambda 5} &=& \frac{1}{4}
\, A_{\rho} F_{\sigma}^{\ \ \rho} F_{\lambda}^{\ \ \sigma} \, .
\end{eqnarray}
\indent
Here the Greek indices $\mu, \nu, \dots $ are raised and lowered by
means of the $4$-dimensional metric tensors $g^{\mu \lambda}$ and
$g_{\lambda \rho}$.

\end{document}